\documentclass[amssymb,useAMS,prd,aps,amsmath,nofootinbib,superscriptaddress,twocolumn]{revtex4}

\usepackage{natbib}
\usepackage{color}
\usepackage{graphicx}
\usepackage{amsmath}
\usepackage[caption=false]{subfig}
\usepackage[colorlinks=true,linkcolor=blue,citecolor=blue,urlcolor=black]{hyperref}

\begin{document}
\title{Spectral and Spatial Distortions of PeV Neutrinos from Scattering with Dark Matter}
\author{Jonathan H. Davis}
\affiliation{Institut d'Astrophysique de Paris, UMR 7095, CNRS, UPMC Universit\'{e} Paris 6, Sorbonne Universit\'{e}s, 98 bis boulevard Arago, 75014 Paris, France}
\author{Joseph Silk}
\affiliation{Institut d'Astrophysique de Paris, UMR 7095, CNRS, UPMC Universit\'{e} Paris 6, Sorbonne Universit\'{e}s, 98 bis boulevard Arago, 75014 Paris, France}
\affiliation{Laboratoire AIM-Paris-Saclay, CEA/DSM/IRFU, CNRS, Universite Paris Diderot,  F-91191 Gif-sur-Yvette, France}
\affiliation{Department of  Physics \& Astronomy, The Johns Hopkins University, \\
3400 N Charles Street, Baltimore, MD 21218, USA}
\affiliation{Beecroft Institute of Particle Astrophysics and Cosmology, Department of Physics,
University of Oxford, Denys Wilkinson Building, 1 Keble Road, Oxford, OX1 3RH, UK
 \\ {\smallskip \tt  \href{mailto:jonathan.h.m.davis@gmail.com}{jonathan.h.m.davis@gmail.com}, \href{mailto:silk@iap.fr}{silk@iap.fr}\smallskip}}

\begin{abstract}
We study the effects on the spectrum and distribution of high-energy neutrinos due to scattering with dark matter both outside and within our galaxy, focusing on the neutrinos observed by the IceCube experiment with energies up to several PeV. If these neutrinos originate from extra-galactic astrophysical sources, then scattering in transit  with dark matter particles will delay their arrival to Earth. This results in a cut-off in their spectrum at an energy set by the scattering cross section, allowing us to place an upper limit on cross sections $\sigma$ which increase with energy $E$ at the level of $\sigma \lesssim 10^{-17} \cdot (m/\mathrm{GeV})\cdot (E / \mathrm{PeV})^2$~cm$^2$, for dark matter particles of mass $m$.
Once these neutrinos enter our galaxy, the large dark matter densities result in further scattering, especially towards the Galactic Centre. Intriguingly, we find that for  $\sigma \sim 10^{-22} \cdot (m/\mathrm{GeV})\cdot (E / \mathrm{PeV})^2$~cm$^2$, the distribution of the  neutrinos on the sky has a small cluster of events towards the centre of the galaxy, potentially explaining the $\sim 2$ sigma excess seen by IceCube in this region without needing a galactic source.
\end{abstract}

\maketitle
\section{Introduction}
Dark matter (DM) is inferred  to exist, through its gravitational interactions with visible matter, within and between galaxies~\cite{Ade:2013zuv,Newman:2012nw,Klypin:2001xu,Iocco:2015xga}. However its nature at the particle level will remain unknown unless its interactions with the particles of the Standard Model are understood. 
One of the least constrained interactions of DM is with neutrinos. Bounds exist on the cross section for scattering at low energy from the Cosmic Microwave Background (CMB) and large-scale structure~\cite{Wilkinson:2014ksa,Bertoni:2014mva,Mangano:2006mp} and limits also exist at MeV energies from supernovae~\cite{Boehm:2013jpa,Mangano:2006mp,Bertoni:2014mva,Farzan:2014gza}.
However the DM-$\nu$ scattering cross section is only weakly constrained at higher neutrino energies (TeV-scale and larger) due to a relative lack of observational tests, both astrophysical and terrestrial. Such high-energy probes are important as they provide information on how DM particles interact at energies inaccessible to colliders, for example.

We show that the recent observation of PeV-energy neutrinos by the IceCube experiment~\cite{Aartsen:2014gkd} allows the DM-$\nu$ scattering cross section to be probed at much higher energies than has previously been possible (see also refs.~\cite{Ioka:2014kca,Kamada:2015era,Borriello:2013ala,Cherry:2014xra,Marfatia:2015hva} for other new physics probes).
Indeed the IceCube experiment has now observed $37$ neutrino events with energies between $30$ and $2000$~TeV after $988$~days of data-taking.
The distribution of these neutrinos is approximately isotropic, implying an extra-galactic origin (see also ref.~\cite{PhysRevD.90.023010}), however there is potentially a small cluster of events towards the Galactic Centre.

In order to study the interactions of these high-energy neutrinos with DM, we consider two different regimes: scattering of neutrinos with DM between galaxies and within the Milky Way itself.
For the former we show in section~\ref{sec:extra} that a large scattering cross section results in a delayed arrival to our galaxy of the neutrinos, since they must diffuse towards Earth instead of free-streaming. Hence the observation of PeV-energy neutrinos by IceCube means that they have not been impeded significantly in transit. 
By making the assumption that the highest energy events observed by IceCube originate from extra-galactic hypernova remnants~\cite{Liu:2013wia,Kelner:2006tc,Chakraborty:2015sta}, we calculate the cross section $\sigma$ for DM-$\nu$ scattering which would result in a cut-off in the observed spectrum below 2~PeV, allowing us to set an upper limit on $\sigma$.

For the case of scattering within our galaxy we show in section~\ref{sec:gal} that DM-$\nu$ scattering results in the generation of an anisotropic distribution of neutrino events from a purely extra-galactic flux, due to enhanced scattering within the Galactic Centre region where the DM density is largest. We calculate the cross section needed to generate the amount of clustering of high-energy neutrino events towards the Galactic Centre potentially observed by the IceCube experiment~\cite{Aartsen:2014gkd}. 
Finally in section~\ref{sec:conc} we summarise our findings and conclude.

\section{Dark Matter-neutrino scattering  in the intergalactic medium \label{sec:extra}}
\begin{figure}[t]
\centering
\includegraphics[width=0.45\textwidth]{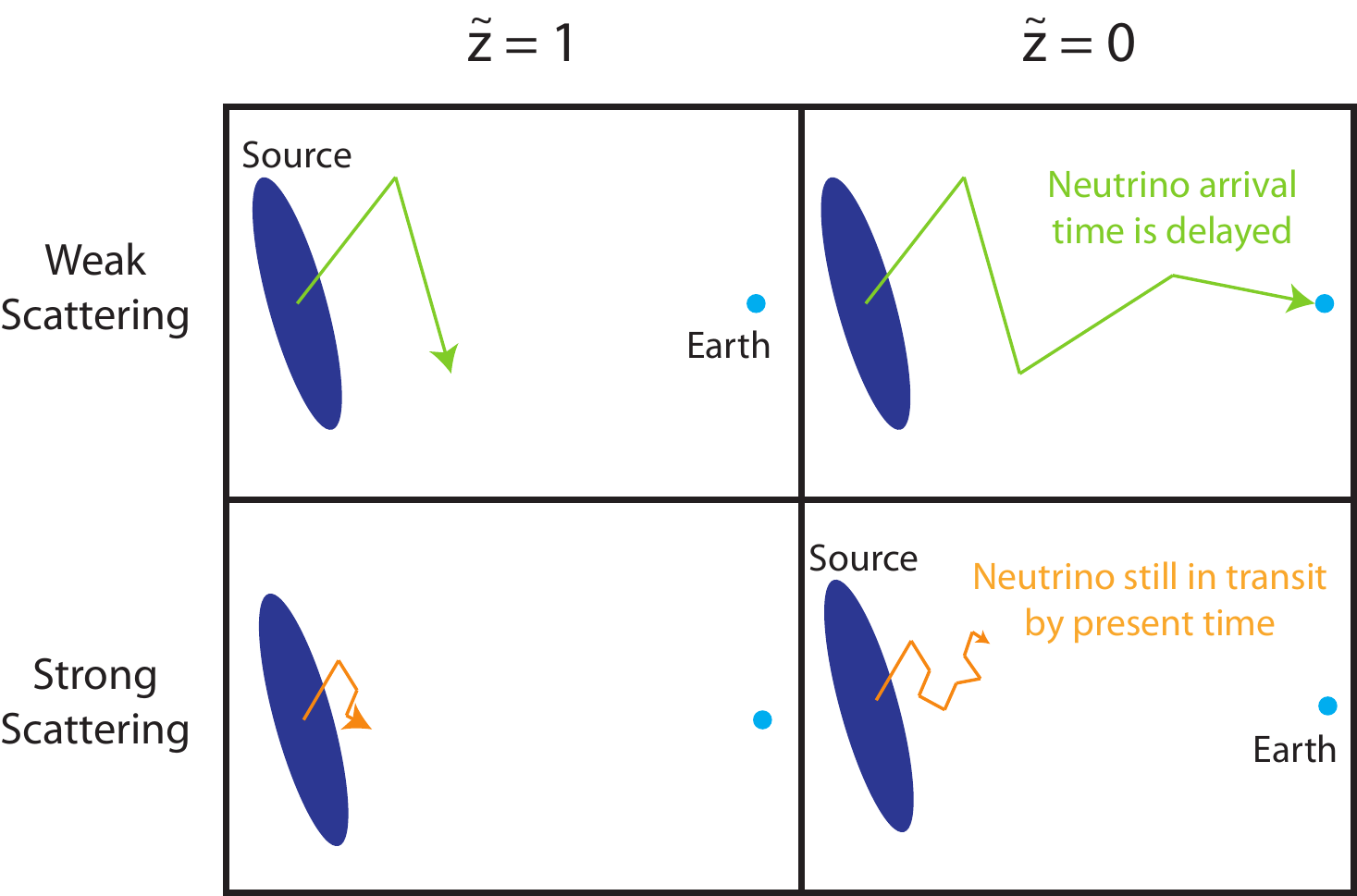}
\caption{Transit of a PeV-energy neutrino from its extra-galactic source to Earth at two redshift values $\tilde{z}$. With weak DM-$\nu$ scattering, the path of the neutrino is lengthened between the  source and Earth, causing it to arrive at the present day $\tilde{z} = 0$. When the scattering rate is large, i.e. strong scattering, the neutrino has to diffuse towards Earth, in which case it has not arrived by the present day.}
\label{fig:nu_scatter_diag}
\end{figure}

Due to their weak interactions with ordinary matter, the PeV neutrinos observed by IceCube should have travelled unimpeded from their extra-galactic sources to Earth. However this may not be the case as the rate at which  these neutrinos scatter with dark matter particles is poorly constrained~\cite{Wilkinson:2014ksa,Bertoni:2014mva,Boehm:2013jpa,Mangano:2006mp} and could be large, resulting in them diffusing towards our galaxy instead of free-streaming.
In this section, we constrain the scattering cross section $\sigma$ for DM between galaxies using the spectrum of high-energy neutrinos observed by IceCube. 

The mean free path for scattering is $\lambda_{\mathrm{DM}-\nu} = m / ( \rho_{\mathrm{DM}} \, \sigma)$, where $m$ is the dark matter mass and $\rho_{\mathrm{DM}}$ is its density in the intergalactic medium (IGM). We assume that the latter takes the form $\rho_{\mathrm{DM}} = \rho_0 (1 + z)^3$, where $\rho_0 = 2.7 \cdot 10^{-27}$~kgm$^{-3}$ and $z$ is the redshift~\cite{Planck:2015xua}.
If $\lambda_{\mathrm{DM}-\nu}$ is smaller than the distance between galaxies, then the neutrinos will scatter multiple times, lengthening their path from the source to Earth and thereby delaying their arrival. We assume that the DM mass is much larger than that of the neutrinos and that the scattering is elastic and flavour-universal~\cite{Mena:2014sja}.

A schematic of this effect is shown in figure~\ref{fig:nu_scatter_diag}, where we give three broad regimes of scattering. 
In the case of `weak scattering', we have that $\sigma > 0$ and so the path the neutrino has to travel is elongated, causing it to arrive later at the present day ($\tilde{z} = 0$). The case of `strong scattering' where $\sigma \gg 0$ has the neutrino scattering so often as it travels from outside our galaxy that it has yet to arrive at Earth by the present day.

We now use this formalism to calculate the effect on the flux of extra-galactic neutrinos.
Without scattering between DM and neutrinos, the expected flux is obtained by integrating over sources at all redshifts $z$~\cite{Liu:2013wia,Kelner:2006tc,Chakraborty:2015sta,Wang:2007ya},
\begin{equation}
\Phi(E){=}{\int} \frac{\mathrm{d} {z} }{4 \pi H_0}    \frac{R({z})}{\sqrt{\Omega_{\mathrm{M}}(1+{z})^3 + \Omega_{\Lambda}}} \, \frac{\mathrm{d}N}{\mathrm{d}E} [(1 + {z})E ] ,
\label{eqn:nu_flux_nodiff}
\end{equation}
where $\mathrm{d}N / \mathrm{d}E$ is the neutrino spectrum at the source, $R(z)$ is the rate of the particular source events for the neutrinos, $H_0 = 0.69$~kms$^{-1}$Mpc$^{-1}$ is the Hubble constant, $\Omega_{\mathrm{M}} = 0.27$ is the fraction of matter in the Universe and $\Omega_{\Lambda} = 0.73$ is the fraction of dark energy~\cite{Planck:2015xua}.
We assume that $R(z)$ is proportional to the star formation rate $R_{\mathrm{SFR}}({z})$~\cite{Liu:2013wia,Kelner:2006tc,Chakraborty:2015sta}.

Since the effect of DM-neutrino scattering is to delay the arrival of the neutrinos to Earth, the redshift at which the neutrino is emitted is no longer necessarily equal to the redshift of the source. Hence the integral over redshift from equation (\ref{eqn:nu_flux_nodiff}) is split into two integrals: one over the redshift distance to the source $z^{\prime}$ and another over the emission time of the neutrinos $\tilde{z}$,
\begin{equation}
\Phi(E){=} {\int}  \frac{\mathrm{d} z^{\prime}}{4 \pi H_0}  {\int}\mathrm{d} \tilde{z} \frac{R(\tilde{z}) f(z^{\prime}, \tilde{z}, \sigma)}{\sqrt{\Omega_{\mathrm{M}}(1+\tilde{z})^3 + \Omega_{\Lambda}}} \, \frac{\mathrm{d}N}{\mathrm{d}E} [(1 + \tilde{z})E ] .
\label{eqn:nu_flux_diff}
\end{equation}

\begin{figure}[b]
\centering
\includegraphics[width=0.48\textwidth]{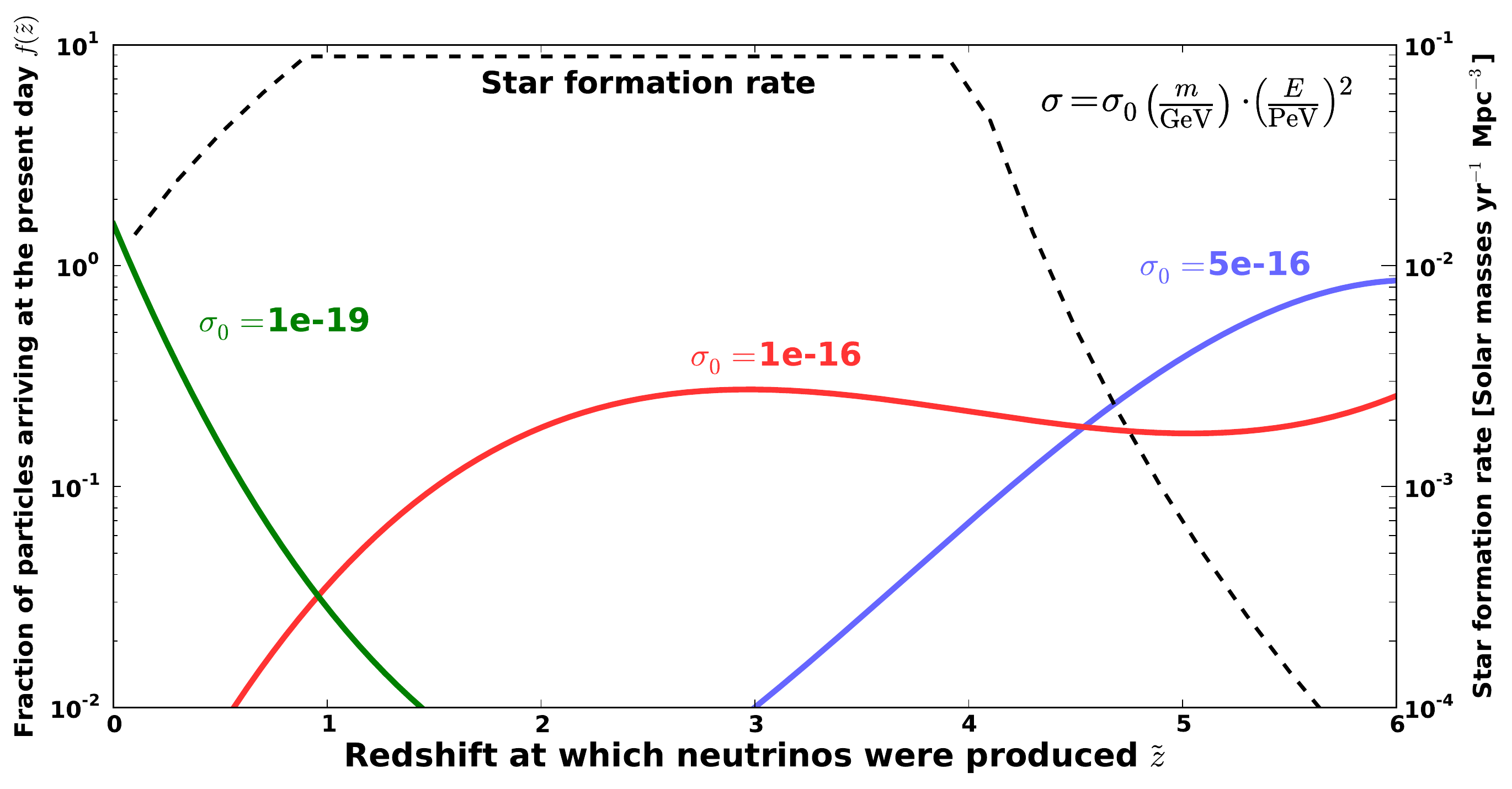}
\caption{Fraction of neutrinos  $f(z^{\prime}, \tilde{z}, \sigma)$ which were produced at a given time in the past represented by a value of redshift $\tilde{z}$, given that they originate from an extra-galactic source at a distance of 10~Mpc from Earth (at the present day). Larger values of the DM-$\nu$ scattering cross section $\sigma$ mean that neutrinos arriving today were produced longer ago. We also show the star formation rate as a function of $\tilde{z}$ for comparison.}
\label{fig:hist_E0=1}
\end{figure}

The time delay due to the DM-neutrino scattering is contained in the (normalised) delay function $f(z^{\prime}, \tilde{z}, \sigma)$, which gives the probability that a neutrino was emitted at a redshift $\tilde{z}$, given that its source is at a distance corresponding to redshift $z^{\prime}$. The amount of delay depends on the cross section of scattering between dark matter and neutrinos $\sigma$, which could itself vary with the neutrino energy $E$.
For figure~\ref{fig:hist_E0=1} and throughout this work we assume that $\sigma = \sigma_0 (m / \mathrm{GeV}) \cdot  \left( {E}/{E_0} \right)^2$, where $E_0 = 1 \, \mathrm{PeV}$.
This is an empirical parameterisation, and to preserve unitarity, the cross section should decrease at higher energies above a few PeV.
Hence in this case $f(z^{\prime}, \tilde{z}, \sigma)$ also depends on the neutrino energy as it travels to Earth.

Examples of $f(z^{\prime}, \tilde{z}, \sigma)$ which have been calculated numerically for different values of $\sigma_0$ are shown in figure~\ref{fig:hist_E0=1}, assuming a source distance of 10~Mpc corresponding to $z^{\prime} \approx  0.0025$  and a neutrino energy at the source of 1~PeV.
For $\sigma_0 = 10^{-19}$~cm$^2$ ,the neutrinos reaching Earth have been produced recently, and so values of $\tilde{z}$ which are close to $z^{\prime}$ will contribute to the integral in equation (\ref{eqn:nu_flux_diff}).
Indeed when $\sigma_0 = 0$ ,we have that $f(z^{\prime}, \tilde{z}, \sigma = 0) \equiv \delta(z^{\prime} - \tilde{z})$, since the neutrino was emitted at the same redshift as we observe the source to be at today and eq. (\ref{eqn:nu_flux_diff}) reduces to eq. (\ref{eqn:nu_flux_nodiff}). 

By contrast for $\sigma_0 = 10^{-16}$~cm$^2$ and $\sigma_0 = 5 \cdot 10^{-16}$~cm$^2$,  the longer path means the neutrinos originated from their source farther in the past. Hence for these cross sections, the integral in equation~(\ref{eqn:nu_flux_diff}) is dominated by values of $\tilde{z}$ much larger than $z^{\prime}$.
As such for $\sigma_0  \gtrsim 5 \cdot 10^{-16}$~cm$^2$, the majority of neutrinos arriving at the present day will have been produced at $\tilde{z} \gtrsim 5$, when the star-formation rate was small.  In this case, the neutrino flux would likely be much smaller than that observed by IceCube.

We now perform the full integral in equation~\ref{eqn:nu_flux_diff} by assuming a specific astrophysical model for the high-energy neutrino flux observed by IceCube. We use the model proposed in refs.~\cite{Liu:2013wia,Kelner:2006tc,Chakraborty:2015sta} in which the highest-energy ($\gtrsim 100$~TeV) neutrinos originate from the decay of hadrons accelerated at extra-galactic hypernova remnants (HNR) (alternative models for these neutrinos use e.g. Active Galactic Nuclei~\cite{Stecker:2013fxa} or gamma-ray bursts~\cite{Tamborra:2015qza}). 
We calculate the expected flux using eqn.~(\ref{eqn:nu_flux_diff}), replacing $R(\tilde{z})$ with the HNR rate which we take to be $R(\tilde{z}) = 10^{-4} M_{\odot}^{-1} R_{\mathrm{SFR}}(\tilde{z})$~\cite{Liu:2013wia,Kelner:2006tc,Chakraborty:2015sta}. For the source neutrino spectrum $\mathrm{d}N / \mathrm{d} E$, we use the formalism of refs.~\cite{Liu:2013wia,Kelner:2006tc,Chakraborty:2015sta}, based on a convolution of the spectrum of protons accelerated at the HNR with the secondary neutrino spectrum produced by each proton (via e.g. pion or muon decay).
This depends on the maximum energy to which these protons can be accelerated, which we take to be $E^p_{\mathrm{max}} = 10^{18}$~eV for the HNR~\cite{Liu:2013wia,Kelner:2006tc,Chakraborty:2015sta}. 

\begin{figure}[t]
\centering
\includegraphics[width=0.49\textwidth]{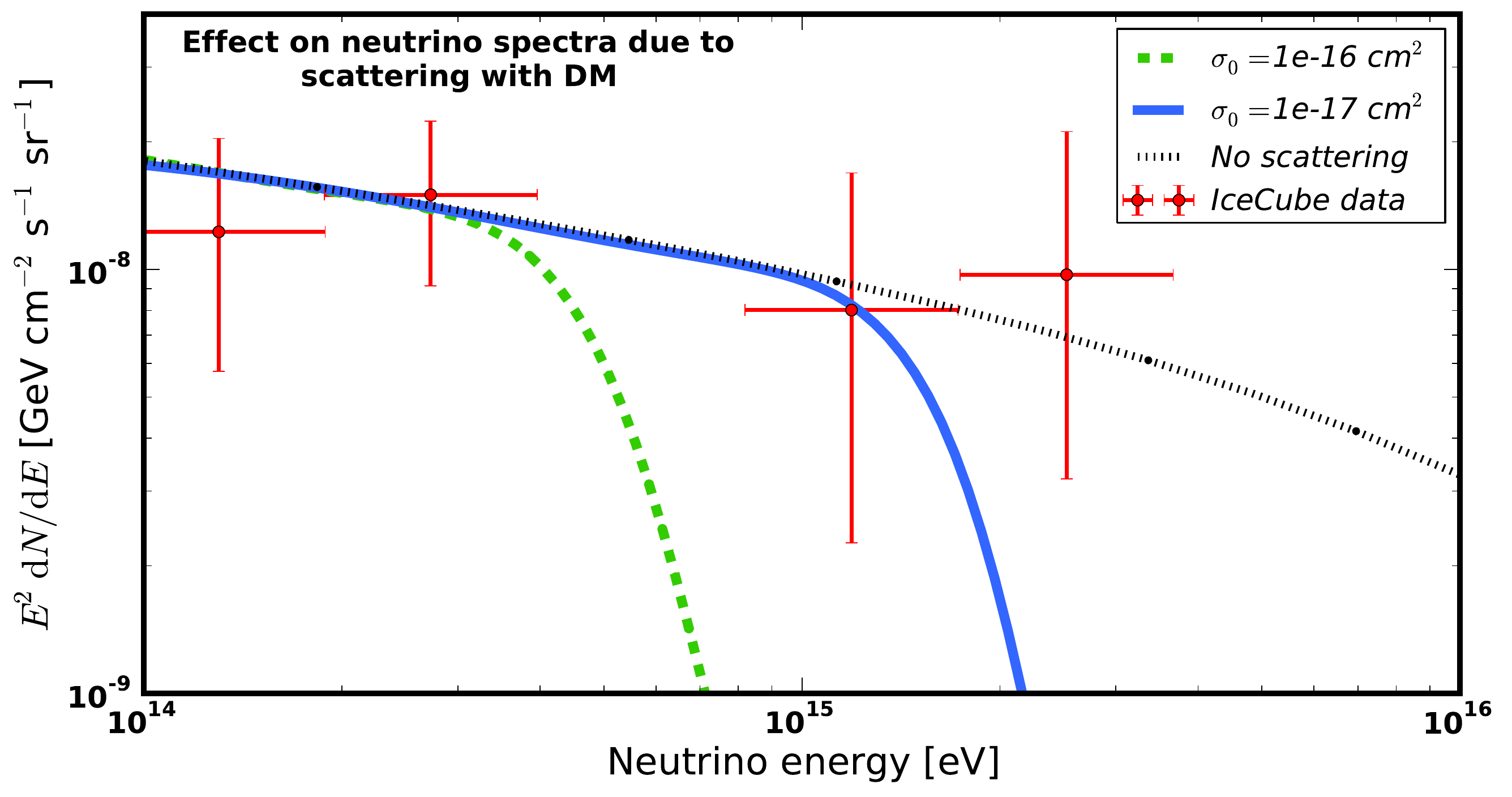}
\caption{Spectra of high-energy neutrinos from extra-galactic hypernova remnants for different values of their scattering cross section with DM. These are compared with the case of zero scattering and data from the IceCube experiment~\cite{Aartsen:2014gkd}. We assume a cross section which increases with neutrino energy $E$ as $\sigma = \sigma_0 (m / \mathrm{GeV}) \cdot  \left( {E}/ {\mathrm{PeV}} \right)^2$. Larger cross sections result in a cut-off in the spectrum at lower energies.}
\label{fig:spectra_dm}
\end{figure}

The effect of scattering between DM and high-energy neutrinos in transit to Earth from their extra-galactic HNR sources is shown in figure~\ref{fig:spectra_dm}, compared with the astrophysical neutrino spectrum observed by IceCube~\cite{Aartsen:2014gkd,Palomares-Ruiz:2015mka}. 
The result is a cut-off in the spectrum of neutrinos at an energy set by the size of $\sigma_0$. This arises from neutrinos which scatter so often that they do not arrive at Earth by the present day, even if they are produced when the first hypernovae form around $\tilde{z} \sim 5$. Larger values of $\sigma_0$ effectively widen the range of energies for which this condition holds. 
Since as can be seen in figure~\ref{fig:spectra_dm} the IceCube data extend up to $E \sim 2$~PeV we conservatively set an upper limit on a DM-$\nu$ cross section which increases with $E^2$ at the level of $\sigma_0 \lesssim 10^{-17}$~cm$^2$. This limit is robust to changes in the model for the production of these neutrinos, as the cut-off is a generic feature of DM-$\nu$ scattering. Future measurements of  higher energy neutrinos will serve to strengthen this upper limit.

\vspace{-10pt}
\section{Scattering within the Milky Way \label{sec:gal}}
\begin{figure*}[t]
\centering
\includegraphics[width=0.46\textwidth]{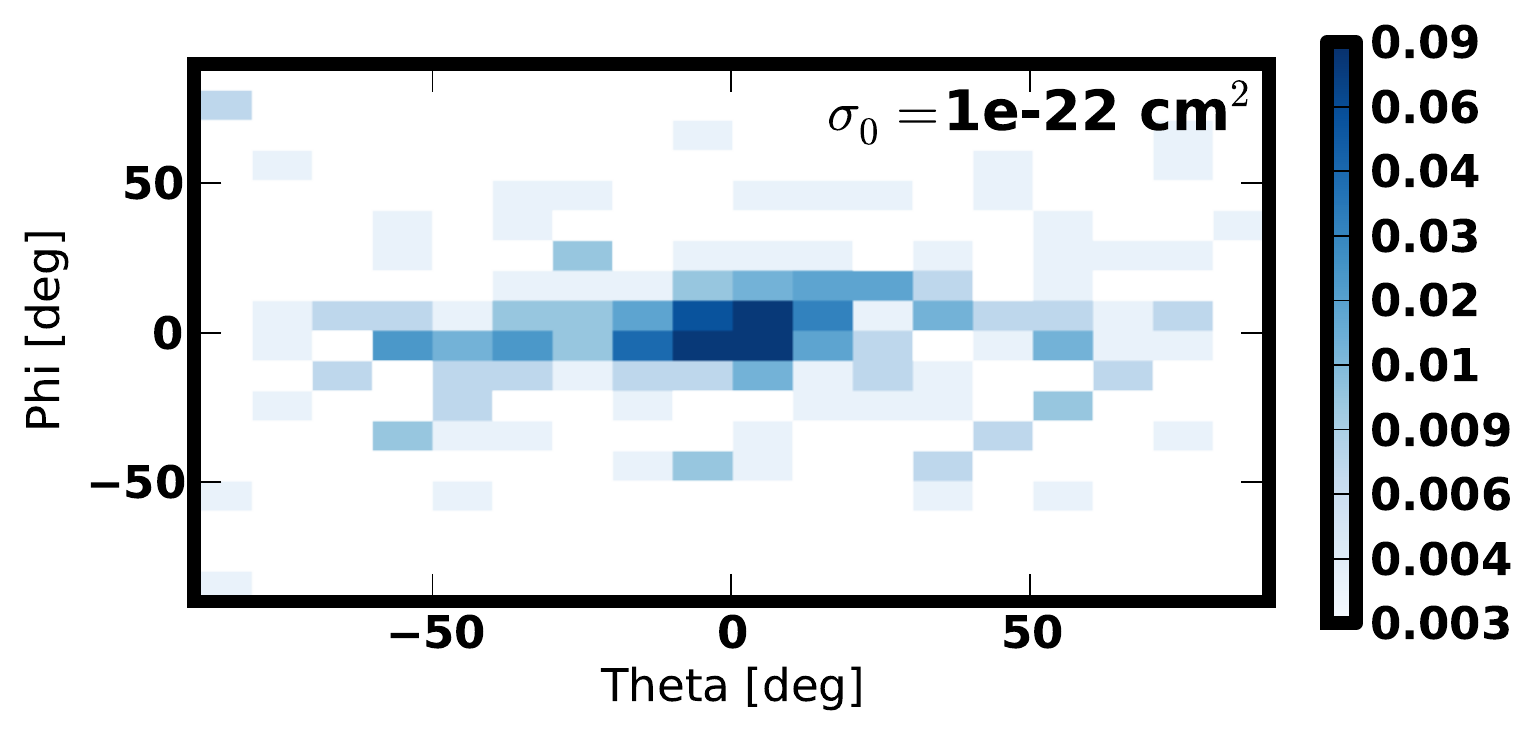} 
\hspace{10pt}
\includegraphics[width=0.46\textwidth]{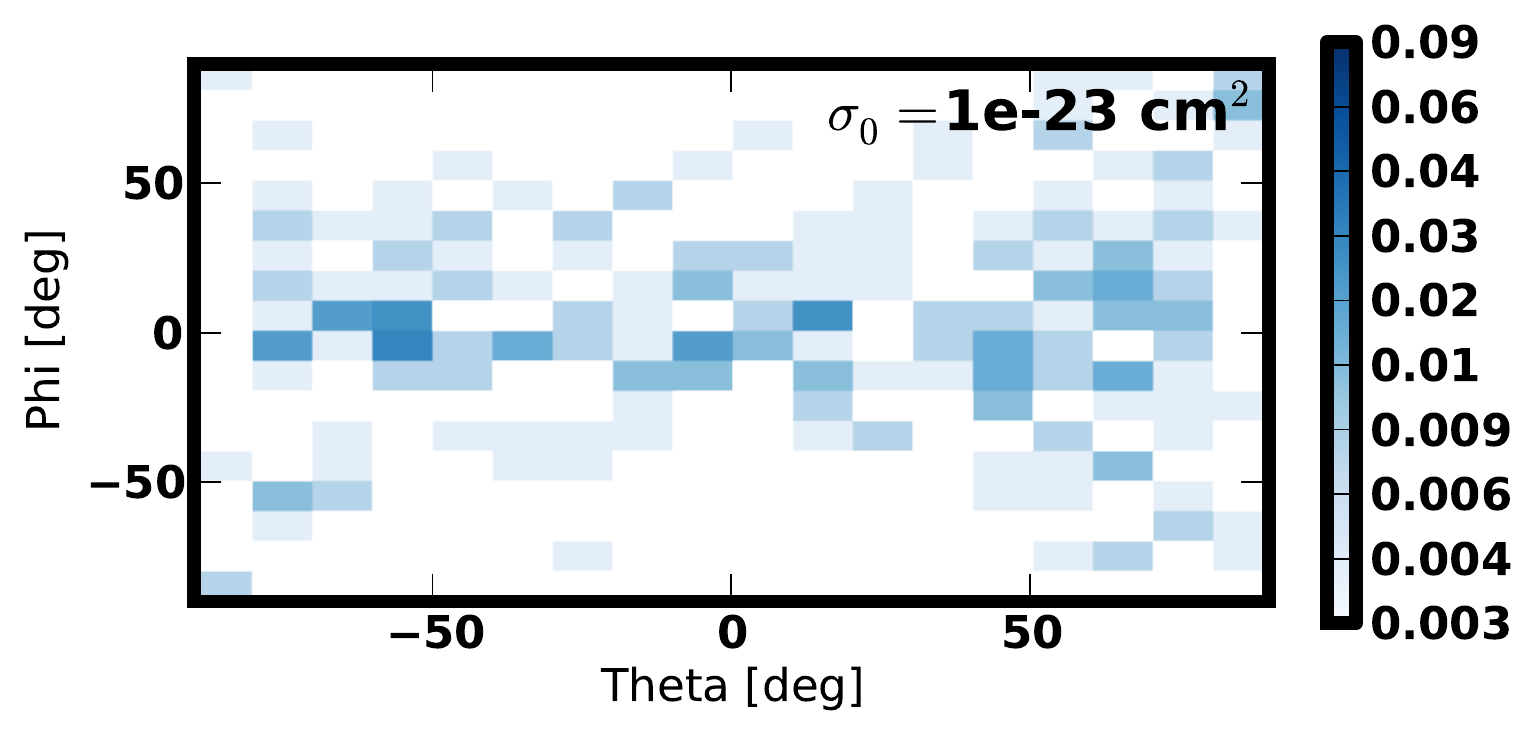} 
\caption{Distribution of neutrinos (with $E = 1$~PeV) on the sky centred on the Galactic Centre for two different values of the scattering cross section with DM  $\sigma = \sigma_0 \cdot (m / \mathrm{GeV})$. The (logarithmic) colour scale gives the fraction of the total rate in each angular pixel. We assume that the neutrinos are produced outside the galaxy and enter isotropically, however for DM-$\nu$ scattering within our galaxy with $\sigma_0 = 10^{-22}$~cm$^2$ the high densities of DM in the Galactic Centre region create a cluster of events. For $\sigma_0 = 10^{-23}$~cm$^2$ by contrast the scattering rate is much lower and so the neutrino flux remains mostly isotropic.}
\label{fig:gc_scatter}
\end{figure*}

In this section, we consider the effects of high-energy extra-galactic neutrinos scattering with DM within our own galaxy, after they have travelled through the IGM. We have already shown that DM-$\nu$ scattering between galaxies  results in a cut-off in the spectrum of these neutrinos for large enough cross sections. However here we show that scattering within the Milky Way, particularly at the centre of the galaxy, affects the distribution of high-energy neutrino events on the sky. This is  interesting as the IceCube collaboration have reported a small ($\lesssim 2\sigma$ significance) clustering of high-energy neutrino events towards the Galactic Centre~\cite{Aartsen:2014gkd,Esmaili:2014rma}, which corresponds to approximately $15\%$ of all of their events.

We assume that the neutrinos enter the galaxy from all directions isotropically as expected if their origin is extra-galactic. However even a kiloparsec from the Galactic Centre, the dark matter density can reach values as large as $\rho \sim 100$~GeVcm$^{-3}$ for a Navarro-Frenk-White (NFW) profile~\cite{Navarro:1996gj}, and the density increases further towards the centre. Hence at this DM density, a value of $\sigma_0 \gtrsim 10^{-23}$~cm$^2$ for a PeV-energy neutrino would be enough for the mean free path $\lambda_{\mathrm{DM}-\nu}$ to be of kiloparsec scale or smaller, and so there can be significant scattering between DM and neutrinos towards the Galactic Centre even given the constraints of the previous section. This will lead to neutrinos appearing to originate from this region even if they are produced outside of our galaxy.

The effect of DM-$\nu$ scattering within our galaxy on the distribution of extra-galactic neutrinos (with an energy of 1~PeV) is shown in figure~\ref{fig:gc_scatter} for two different values of the scattering cross section (assuming an NFW profile for the DM). For $\sigma_0 = 10^{-22}$~cm$^2$, the scattering between DM and PeV-energy neutrinos generates a concentration of events towards the Galactic Centre, where the DM density (and therefore the scattering rate) is largest, at the level of about $15 - 20\%$ of the total rate. In this case scattering has altered the direction of travel for these neutrinos and therefore their apparent point of origin.
By contrast for $\sigma_0 = 10^{-23}$~cm$^2$ (and below), the neutrino flux remains mostly isotropic. 
Hence $\sigma_0 = 10^{-22}$~cm$^2$ gives a similar amount of clustering towards the Galactic Centre seen in the latest IceCube data~\cite{Aartsen:2014gkd,Esmaili:2014rma}, while values larger than this would introduce too much anisotropy.

The Galactic Centre is not the only region for which an anisotropy could be observed: the large DM densities towards the cores of dwarf galaxies~\cite{Mashchenko:2015qda,Governato:2009bg,Walker:2009zp} could result in a similar effect from DM-$\nu$ scattering. The size of this anisotropy depends on the mass model for the dwarf galaxies, which is not well known. For models where the DM density is $\gtrsim 100$~GeVcm$^{-3}$ within 1~kpc of the  centre~\cite{Governato:2009bg}, the mean free path with $\sigma_0 = 10^{-22}$~cm$^2$ is smaller than 1~kpc. Hence the flux of high energy neutrinos around e.g. the Fornax or Sculptor dwarfs could be enhanced by as much as a few percent compared to the average all-sky flux, depending on the density profile.

\vspace{-10pt}
\section{Conclusion \label{sec:conc}}
\begin{table}[b]
\begin{center}
\begin{tabular}{ c || c | c | c }
 & \normalsize{$\sigma  (m / \mathrm{GeV})^{-1}$ [cm$^2$]} & Energy-scale \\
\hline
 \normalsize{This work (gal.)} & \normalsize{$\sim 10^{-52} \cdot (E / \mathrm{eV})^2$} & \normalsize{100 - 2000~TeV}  \\
\normalsize{This work (ex-gal)} & \normalsize{$\lesssim 10^{-47}\cdot (E / \mathrm{eV})^2$} & \normalsize{100 - 2000~TeV} \\
  \normalsize{Structure form.~\cite{Wilkinson:2014ksa}} & \normalsize{$\lesssim 10^{-38}  \cdot (E / \mathrm{eV})^2$} & \normalsize{$\lesssim$~eV}  \\
    \normalsize{SN 1987A~\cite{Boehm:2013jpa}} & \normalsize{$\lesssim 10^{-28}$} & \normalsize{$\sim 10$~MeV}  \\
\end{tabular}
\end{center}
\caption{Here `gal.' refers to best-fit cross section $\sigma$ to give the required level of clustering seen towards the Galactic Centre in the recent IceCube data~\cite{Aartsen:2014gkd}. `Ex-gal' is our upper limit from requiring that neutrinos with energies of 2~PeV and below are able to reach Earth by the present day.}
\label{table_constraints}
\end{table}

We have investigated the effects of scattering between high-energy neutrinos and dark matter particles, focusing on the neutrino events observed by the IceCube experiment between energies of $E \approx 30 - 2000$~TeV~\cite{Aartsen:2014gkd}. Since there are strong arguments for an extra-galactic origin of these neutrinos~\cite{Aartsen:2014gkd,PhysRevD.90.023010} we have studied the effect of scattering both outside and within our galaxy. 
For extra-galactic DM-$\nu$ scattering with a cross section which rises with $E^2$, we showed that this results in a delayed arrival for the neutrinos to Earth (figures~\ref{fig:nu_scatter_diag} and~\ref{fig:hist_E0=1}), due to the large distances ($\gtrsim 10$~Mpc) they must diffuse through when the scattering rate is large. This leads to a cut-off in their \emph{spectrum} at an energy set by the size of the cross section (figure~\ref{fig:spectra_dm}). While for scattering within our galaxy, we found that the \emph{distribution} of the neutrinos is shifted from pure isotropy to possessing an excess of events towards the Galactic Centre, where the DM density is largest (figure~\ref{fig:gc_scatter}). This results from scattering altering the direction of travel for these neutrinos, making them appear to originate from the Galactic Centre.

We summarise our numerical results compared with those from previous works in table~\ref{table_constraints}. As can be seen, our results probe much higher energy scales than previous constraints from structure formation~\cite{Wilkinson:2014ksa,Bertoni:2014mva} and supernova 1987A~\cite{Boehm:2013jpa,Bertoni:2014mva,Mangano:2006mp}.
Throughout this work, we have assumed that the scattering cross section $\sigma$ increases as $E^2$ up to a few PeV (or above), since in this case our constraint is particularly strong compared to those set at lower energies. 
However this is not the only form which the DM-$\nu$ scattering cross section can take, and the complimentarily between the various constraints serves as an important probe of the energy dependence of $\sigma$.
The upper bound labelled `ex-gal' in table~\ref{table_constraints} has been set by assuming that the neutrinos originate from hypernova remnants in other galaxies, although any other sources for these neutrinos, astrophysical~\cite{Stecker:2013fxa,Tamborra:2015qza} or otherwise~\cite{Feldstein:2013kka,Murase:2015gea,Esmaili:2013gha,Zavala:2014dla,Esmaili:2014rma}, would lead to similar constraints. 
For cross sections larger than this upper limit, neutrinos with energies below 2~PeV are impeded too much on their way to our galaxy, which would lead to a cut-off in the IceCube neutrino spectrum at this energy. 
Scattering cross sections just below this bound give no observable cut-off while also leading to interesting phenomenology within our galaxy.
Indeed the best-fit cross section to give a small cluster of PeV-energy events ($\sim 15 - 20\%$) towards the Galactic Centre, due to DM-$\nu$ scattering within our galaxy, is $\sim 10^{-22} \cdot (m / \mathrm{GeV})$~cm$^2$ (labelled `gal' in table~\ref{table_constraints}).
This is consistent with the $\lesssim 2\sigma$ anisotropy observed in the IceCube data-set in this region~\cite{Aartsen:2014gkd,Esmaili:2014rma}.
Cross-sections of this size which scale with $E^2$ also evade the low-energy constraints in table~\ref{table_constraints}.

Currently there are not enough statistics to make a more precise statement. However we predict that if DM particles scatter with  $\sim$PeV energy neutrinos with a cross section around $10^{-22} \cdot (m / \mathrm{GeV})$~cm$^2$, then future data from IceCube (and experiments such as KM3NeT~\cite{Margiotta:2014eaa} or \textsc{Antares}~\cite{Hernandez-Rey:2014ssa}) should show neutrinos with a mostly isotropic distribution plus an excess of events ($\sim 15\%$ of the population) towards the Galactic Centre. The neutrinos are produced in extra-galactic astrophysical sources, with the DM only contributing to the anisotropy through scattering, and so the spectrum will retain the general power-law form expected from such sources {regardless of position on the sky} (although the exact spectrum depends on the energy dependence of the cross section). This combined spectral and spatial dependence is a potentially unique prediction compared to alternative models for the anisotropy.
For example although an astrophysical source within our galaxy (on top of an extra-galactic component) may result in a similar prediction, it would likely over-produce other signals (e.g. $\gamma$ or cosmic rays)~\cite{Joshi:2013aua,PhysRevD.90.023010}. Additionally results from \textsc{Antares} disfavour a point source~\cite{Hernandez-Rey:2014ssa}. Furthermore models where the neutrinos are produced by decaying or annihilating DM lead to an excess of neutrinos towards the Galactic Centre, but the {spectrum} of these events will be very different from those away from this region~\cite{Feldstein:2013kka,Murase:2015gea,Esmaili:2013gha,Zavala:2014dla,Esmaili:2014rma}.
We also expect similar $\sim 10\%$ anisotropies towards other galaxies such as M87, within the regions they subtend on the sky, and potentially small anisotropies for the dwarf galaxies near the Milky Way.
Our results are complimentary to bounds on DM-$\nu$ interactions at lower energies~\cite{Wilkinson:2014ksa,Bertoni:2014mva,Mangano:2006mp,Boehm:2013jpa} and also to probes of scattering of e.g. DM with quarks~\cite{Akerib:2013tjd,Aprile:2012_225,Agnese:2014aze}, DM with electrons~\cite{Essig:2012yx}, DM with photons~\cite{Davis:2014zda,Wilkinson:2013kia} and self-interactions~\cite{Harvey:2015hha}, placing us closer than ever to understanding the nature of~DM.

\vspace{-20pt}
\section*{Acknowledgements}
The research of JHD and JS has been supported at IAP by  ERC project 267117 (DARK) hosted by Universit\'e Pierre et Marie Curie - Paris 6 and also for JS at JHU by National Science Foundation grant OIA-1124403.

\end{document}